\documentclass[prl,reprint,amsmath,amssymb,aps,longbibliography]{revtex4-2}

\usepackage{graphicx}
\usepackage{dcolumn}
\usepackage{bm}
\usepackage{booktabs}

\begin{document}

\preprint{APS/123-QED}

\title{The Nodal Line of the Galaxy Correlation Function as a Geometric Cosmological Probe}

\author{Lado Samushia}
\affiliation{Department of Physics, Kansas State University, Manhattan, Kansas 66506, USA}
 \affiliation{Georgian National Astrophysical Observatory, GE-0179, Georgia.}
 
\date{\today}

\begin{abstract}
We propose a geometric cosmological probe: the nodal line of the anisotropic galaxy correlation function --- the locus in the plane where the function changes sign. Because the nodal line depends only on this sign, it is invariant under any amplitude rescaling, and its information is mostly independent of and complementary to BAO. In a proof-of-concept analysis of DESI DR1 galaxies, the nodal line constrains $\Omega_m h^{3/2}$ to 3\% at fixed galaxy bias, degrading to 4.8\% under a 10\% prior. Combined with BAO, it yields a 4.45\% sound-horizon-free constraint on $H_0$.
\end{abstract}

\maketitle

\textit{Introduction}---The anisotropic galaxy correlation function, $\xi$, is an extremely rich source of cosmological information. Its measured values as functions of the galaxy separations along and transverse to the line of sight (LOS), $\pi$ and $\sigma$, respectively, depend on both cosmological parameters and galaxy properties and have been used to constrain both. In real space, $\xi(\sigma,\pi)$ has two features with model-independent geometric meanings: the Baryon Acoustic Oscillation (BAO) peak at roughly $100\,h^{-1}\,\mathrm{Mpc}$ \cite{Eisenstein_1998, Eisenstein_2005, Cole_2005} and the zero crossing (ZC) at a slightly larger separation \cite{labini2005zerocrossingscaleproblemgalaxy, prada2011measuringequalityhorizonzerocrossing}. A related geometric compression of the BAO feature is the linear point, defined as the midpoint between the BAO peak and the preceding dip \cite{Anselmi_2015, Anselmi_2018a, Anselmi_2018b}. It is worth noting that, although the BAO feature itself has a model-independent geometric interpretation, current measurement practice uses templates that can, in principle, introduce mild model dependence \cite{Vargas_Maga_a_2018, Anselmi_2019, Bernal_2020, Chen_2024, P_rez_Fern_ndez_2025}. While modeling the full 2D shape of $\xi(\sigma,\pi)$ contains, in principle, all of the available information, it is susceptible to various observational and theoretical systematic effects. Small offsets in the predicted value of $\xi$ can accumulate and result in strongly biased inferences \cite{Findlay_2025, Gambardella_2024, rosadomarin2025mitigatingimagingsystematicsdesi}. Purely geometric probes are less sensitive to this problem.

\begin{figure*}
\includegraphics{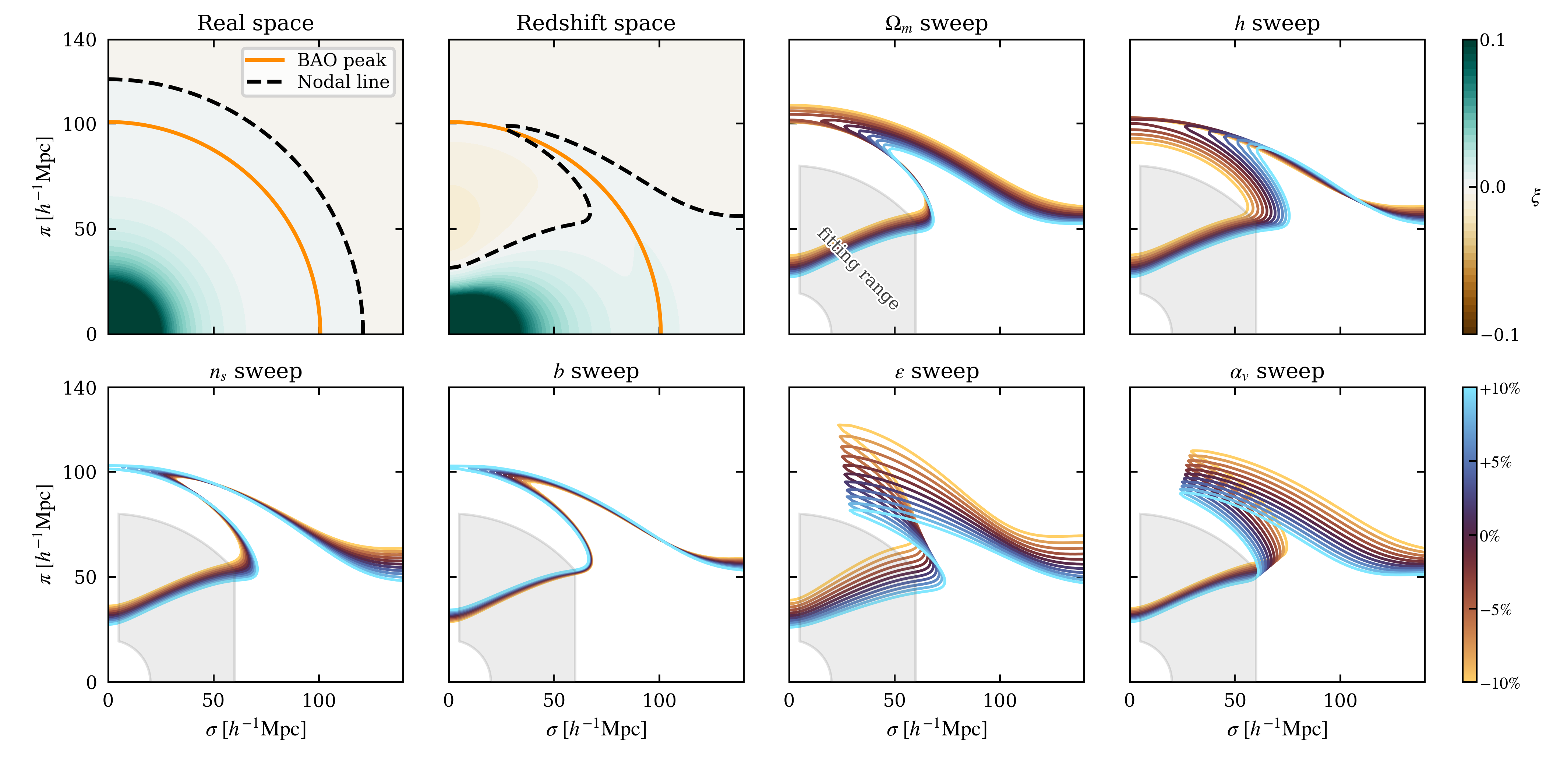}
\caption{\label{fig:motivation} Top two leftmost panels show linear theory predictions for a correlation function in real space and redshift space. Remaining panels show the dependence of the position of zero crossing and the nodal line on cosmological parameters.}
\end{figure*}

To leading order, both the BAO and ZC are robust to multiplicative systematic effects, while the BAO is also robust to additive systematic effects. Peculiar velocities, however, strongly distort these real-space signals. Their effect on the BAO is relatively small, but the ZC is distorted into a fin-like shape that is no longer confined to a special separation and instead crosses multiple scales, as shown in Fig.~\ref{fig:motivation}. BAO is currently considered one of the cleanest and most constraining cosmological probes and has provided an interesting hint of phantom dark energy \cite{Abdul_Karim_2025}. The zero crossing of the angle-averaged correlation function, $\xi_0$, has been proposed as a standard ruler associated with the matter--radiation equality scale, but robust inference is challenging because of sample variance and finite-volume effects.

The main insight of this paper is that, if one wants to use the ZC as a cosmological probe, measuring it in $\xi_0$ is suboptimal. The function $\xi_0$ will eventually cross zero on sufficiently large scales, where the negative-$\xi$ regions in Fig.~\ref{fig:motivation} overwhelm the positive-$\xi$ regions. However, it is clear that detecting the ZC in this way is not optimal and does not respect its geometry. We propose instead that the ZC is best detected in the interval between 20 and $80\,h^{-1}\,\mathrm{Mpc}$, indicated by the gray region in Fig.~\ref{fig:motivation}. The ZC in this region is nearly a straight line and can be described by two numbers: its angle with respect to the horizontal axis, $\phi$, and its perpendicular offset from the origin, $c$. We refer to the segment of the ZC in this region as the \textit{nodal line} (NL).

The NL has several advantages over the ZC in $\xi_0$. The correlation function has a much steeper gradient and larger surrounding amplitude near the NL than near the large-scale zero crossing of $\xi_0$; this makes the NL easier to detect and less sensitive to additive systematics whose effect scales inversely with the derivative \cite{de_Mattia_2019}. The NL occupies scales distinct from the BAO and is therefore expected to provide largely complementary information. It can also be made insensitive to the small-separation and small-transverse-separation regions that are most susceptible to theoretical and observational systematic effects \cite{bianchi2025characterizationdesifiberassignment, Pinon_2025}.

\textit{Physics of the Nodal Line}---We define the NL as the portion of the ZC satisfying
\[
20 < \sqrt{\sigma^2+\pi^2} < 80\,h^{-1}\,\mathrm{Mpc},
\qquad
5 < \sigma < 60\,h^{-1}\,\mathrm{Mpc}.
\]
The two upper-right panels and the four bottom panels of Fig.~\ref{fig:motivation} show how the NL changes in linear theory \cite{1987MNRAS.227....1K, Hamilton:1992zz, Ballinger_1996, Hamilton_1998} as the cosmological parameters and linear bias $b$ are varied. The parameters $\Omega_\mathrm{m}$, $h$, $n_s$, and $b$ primarily affect $c$. The first three do so by changing the broadband shape of the matter power spectrum and the matter--radiation equality scale that sets the real-space ZC, whereas $b$ changes the strength of redshift-space distortions, which are larger for lower bias at fixed cosmology. The weaker sensitivity of the NL to $b$ than to the cosmological parameters is an additional advantage. The Alcock--Paczy\'nski (AP) parameters \cite{1979Natur.281..358A, Matsubara_1996, Padmanabhan_2008, 2011MNRAS.410.1993S} $\varepsilon$, which controls anisotropic stretching, and $\alpha_\mathrm{V}$, which controls isotropic dilation, produce distinct signatures. They affect both $\phi$ and $c$, with the dependence on $\varepsilon$ being especially strong. The modeling can be extended beyond linear theory using perturbative approaches or simulations.

\textit{Detecting the Nodal Line}---We measure $\xi(\sigma,\pi)$ in 1$\,h^{-1}\,\mathrm{Mpc}$ bins and reduce it to its sign field, $q_i=\operatorname{sgn}[\xi(\sigma_i,\pi_i)]$, retaining the location of the zero crossing while discarding all sensitivity to the amplitude of $\xi$ (see Appendix A). Within the NL mask we fit a straight line in Hesse normal form---angle $\phi$ to the $\sigma$ axis, signed perpendicular offset $c$ from the origin---to $q_i$, minimizing the summed squared mismatch against a $\tanh$-softened model sign (softening scale $\Delta=2\,h^{-1}\,\mathrm{mpc}$). Minimization proceeds on a grid in $(\phi,c)$ with a local paraboloid refinement, yielding best-fit values $(\phi_\star,c_\star)$ (see Appendix B). The estimator is deliberately simple and is not statistically optimal: because neighboring $\xi$ bins are correlated and the reliability of their measured signs varies across the plane, an optimal estimator would incorporate the covariance of the sign field or bin-dependent weights. We leave the construction of such an estimator to future work.

\begin{figure*}
\includegraphics{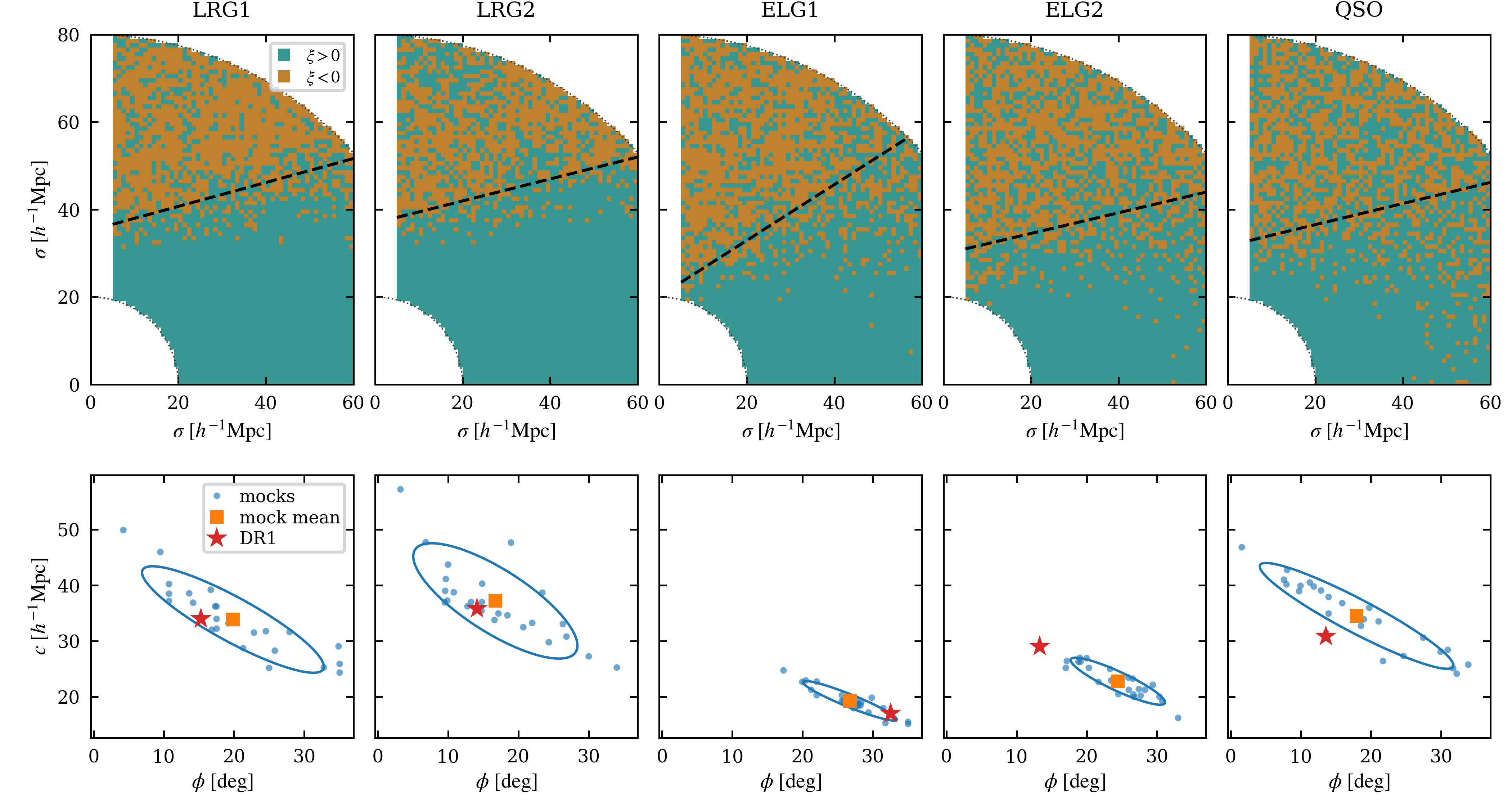}
\caption{\label{fig:dr1} The top row shows the sign of the measured correlation
function in five DESI DR1 redshift bins, together with the best-fit
nodal line shown dashed. The bottom row shows the best-fit angle
$\phi_\star$ and offset $c_\star$ measured from DR1, the corresponding
measurements from the mocks, their mean, and the implied $68\%$
Gaussian uncertainty for the DR1 measurement.}
\end{figure*}

\textit{DESI DR1 Measurements}---We use the official DESI DR1 pipeline and the publicly available DR1 galaxy and mock catalogs to measure the two-dimensional galaxy correlation function over the redshift range $0.4<z<2.1$ \cite{Adame_2025, desicollaboration2026datarelease1dark, ross2024constructionlargescalestructurecatalogs}. We divide the data into the redshift bins used in the official DESI DR1 analysis: LRG1, $0.4<z<0.6$; LRG2, $0.6<z<0.8$; ELG1, $0.8<z<1.1$; ELG2, $1.1<z<1.6$; and QSO, $1.6<z<2.1$ \cite{Adame_2025b}. We omit the BGS bin because it is too noisy to yield a reliable NL measurement and does not contribute significantly to the final constraints. We also do not combine the LRG and ELG tracers in their overlapping redshift range, as doing so would complicate the interpretation of the bias parameter.

The measured correlation functions are shown in the top row of Fig.~\ref{fig:dr1}. The bottom row shows the best-fit $\phi_\star$ and $c_\star$ values for all redshift bins, with those obtained from 25 DR1 mocks per tracer that reproduce the corresponding tracer bias and survey geometry. Combining all five tracers, the DESI DR1 measurements are consistent with the mock distribution, with $\chi^2=16.8$ for 10 degrees of freedom ($p=0.08$), two per redshift bin. The largest individual contribution to the total $\chi^2$ comes from the QSO bin.

The measurement covariance of $(\phi_\star,c_\star)$ is estimated from the per-tracer mock scatter under a two-parameter Gaussian approximation and enters both the consistency test above and the likelihood below (see Appendix C). The parameters are strongly anticorrelated, with larger angles corresponding to offsets closer to the origin. The ELG bins have lower $c_\star$, as expected from their lower galaxy bias.

\textit{Cosmological Interpretation}---For a proof-of-concept inference we predict $\xi$ from linear theory in flat $\Lambda$CDM, apply the same NL-detection algorithm used on the data, and build a Gaussian likelihood $\mathcal{L}(\Omega_\mathrm{m},h,\boldsymbol{b})$ from the DR1 measurements and the mock covariance. We vary $\Omega_\mathrm{m}$, $h$, and one linear bias per tracer (collectively $\boldsymbol{b}$), fixing $n_s$, whose effect on the NL is negligible over the range allowed by the Cosmic Microwave Background (CMB) data.  $\Omega_\mathrm{m}$ and $h$ enter through the intrinsic shape of $\xi$ and through the AP parameters $\varepsilon(\Omega_\mathrm{m})$ and $\alpha_\mathrm{V}(\Omega_\mathrm{m})$.

To gauge sensitivity to galaxy bias, we run the inference with the tracer biases fixed to fiducial values, and again with independent 10\% Gaussian priors on them (fiducial values and their derivation in Appendix D). This suffices for a proof of concept, though a complete analysis would treat bias more self-consistently.

\begin{figure}
\includegraphics{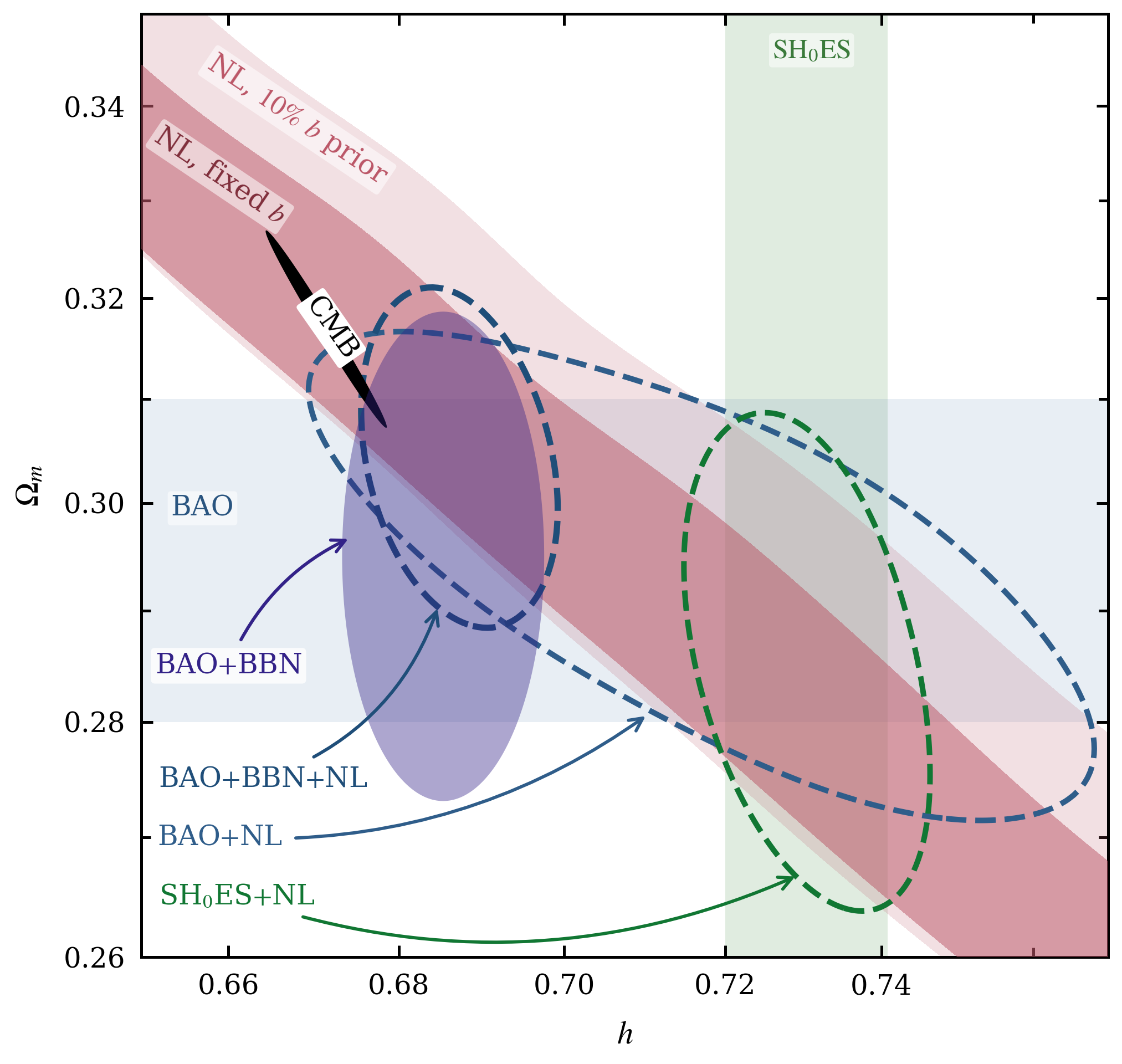}
\caption{\label{fig:cosmo} One-$\sigma$ confidence contours on $\Lambda$CDM parameters from SH$_0$ES (green), DESI BAO (blue), NL with fixed bias and a ten percent bias prior (shades of red), CMB (black), BAO $+$ BBN (indigo). The effect of combining with NL is shown by dashed contours of matching color.}
\end{figure}

The resulting 68\% credible regions are shown in Fig.~\ref{fig:cosmo}, together with constraints from DESI DR1 BAO \cite{Adame_2025c, Adame_2025d}, DESI DR1 BAO combined with a Big Bang nucleosynthesis prior (BBN) \cite{schoneberg20262024bbnbaryonabundance}, Planck CMB observations \cite{Planck_2020}, and the local SH$_0$ES measurement \cite{Riess_2022}. The NL measurement primarily constrains the approximate parameter combination $\Omega_\mathrm{m}h^{3/2}$, approximately the same combination as full shape fits \cite{2023JCAP...04..023B}. The NL-only constraint is $\Omega_\mathrm{m}h^{3/2}=0.1761\pm0.0053$ with the bias parameters fixed and $\Omega_\mathrm{m}h^{3/2}=0.1795\pm0.0087$ under 10\% Gaussian bias priors; marginalizing over bias broadens the posterior asymmetrically toward lower bias, and the latter value is a symmetric summary of that posterior (see Appendix E).

Table~\ref{tab:constraints} illustrates the improvement in parameter constraints obtained by including the NL measurement. The NL constrains $\Omega_\mathrm{m}h^{3/2}$ without using the sound horizon as a calibrated standard ruler, making it a feature-based geometric complement to standard distance probes. When combined with DESI DR1 BAO, the NL determines $h$, and hence $H_0$, to 4.45\% using galaxy clustering alone. This determination requires no calibration of the sound horizon: after marginalizing over $r_\mathrm{d}$, the redshift dependence of the BAO distance ratios constrains $\Omega_\mathrm{m}$, while the NL independently constrains $\Omega_\mathrm{m}h^{3/2}$. Their intersection therefore determines $h$ without a BBN prior or a CMB-calibrated value of $r_\mathrm{d}$. This provides a sound-horizon-independent determination of $H_0$ from clustering geometry, without fitting the full broadband shape or amplitude of $\xi$.

Sound-horizon-independent determinations of $H_0$ have also been obtained from broadband galaxy clustering, either through full-shape information associated primarily with matter--radiation equality or through direct measurements of the power-spectrum turnover \cite{Philcox_2021, Farren_2022, zaborowski2025soundhorizonfreemeasurementh0, Bahr_Kalus_2023, Bahr_Kalus_2025}. The NL provides a methodologically independent and complementary approach, as it uses the geometry of the zero contour rather than fitting the full broadband shape or directly measuring the turnover scale. Although the NL is influenced by the broadband shape and matter--radiation equality physics, it does not treat the equality horizon as an externally calibrated standard ruler; instead, it compresses this information into the position and orientation of the zero contour.

The same complementarity runs the other way: combined with the local SH$_0$ES determination of $h$, the NL constrains $\Omega_\mathrm{m}$ to 5.24\% without using BAO or CMB distance information, and adding it to BAO$+$BBN sharpens $\Omega_\mathrm{m}$ from 5.07\% to 3.61\% (Table~\ref{tab:constraints}). We emphasize that this is a proof-of-concept analysis rather than a final cosmological measurement. Although the quoted uncertainties illustrate the potential constraining power of the NL, the central values should not be interpreted as definitive cosmological constraints because of the simplified linear-theory model and treatment of galaxy bias.

\begin{table}
\caption{
Marginalized posterior means and symmetric $1\sigma$ uncertainties.
The NL combinations use a 10\% Gaussian prior on the tracer bias
parameters.
}
\label{tab:constraints}
\centering
\begin{tabular}{@{}lcc@{}}
\toprule
Data set & $\Omega_{\rm m}$ & $h$ \\
\midrule
BAO
& $0.295 \pm 0.015$
& --- \\
BAO$+$NL
& $0.294 \pm 0.015$
& $0.719 \pm 0.032$ \\

\addlinespace[5pt]
BAO$+$BBN
& $0.296 \pm 0.015$
& $0.685 \pm 0.008$ \\
BAO$+$BBN$+$NL
& $0.305 \pm 0.011$
& $0.687 \pm 0.008$ \\

\addlinespace[5pt]
SH$_0$ES
& ---
& $0.730 \pm 0.010$ \\
SH$_0$ES$+$NL
& $0.286 \pm 0.015$
& $0.730 \pm 0.010$ \\
\bottomrule
\end{tabular}
\end{table}

\textit{Bias Treatment}---Any analysis of galaxy clustering must account for galaxy bias. We expect bias modeling for the NL to be simpler than for analyses of the full shape of $\xi$. Because the NL estimator uses the location of the zero contour rather than the amplitude of the correlation function, it may not require a detailed bias expansion describing the dependence of the galaxy field on the matter density, tidal field, and higher-derivative operators \cite{McDonald_2009, Assassi_2014, Desjacques_2018}. Instead, its leading dependence on galaxy bias enters through the redshift-space anisotropy and may be captured by one effective tracer-bias parameter over the fitted range of scales, possibly supplemented by an additional parameter describing residual scale dependence.

In our proof-of-concept analysis, we adopt independent 10\% Gaussian priors on the tracer biases. These priors are conservative relative to the approximately 1--3\% and 2--3\% constraints obtained for the LRG and ELG samples, respectively, from small-scale clustering in the DESI One-Percent Survey, and are comparable to the approximately 10--15\% constraint obtained for the QSO sample \cite{yuan2023desionepercentsurveyexploring, rocher2024desionepercentsurveyexploring}. Analyses using larger DESI samples and improved small-scale clustering models are expected to tighten these constraints further. By restricting the NL fit to separations above $20\,h^{-1}\,\mathrm{Mpc}$, we limit its direct overlap with the nonlinear scales that provide most of the information in these bias measurements, which should reduce the covariance between the NL observable and the external bias constraints. The adequacy of this reduced bias description will need to be validated against simulations spanning realistic variations in halo occupation, velocity bias, and tracer properties.

\textit{Conclusions}---We have presented a proof-of-concept extraction of cosmological constraints from the nodal line of the galaxy correlation function. The proposal is conceptually related to analyses of the zero crossing in the angle-averaged correlation function $\xi_0$, but differs substantially in implementation. It is also distinct from tomographic AP tests, which extract the signal from the modeled redshift evolution of the full small-scale $\xi(\sigma,\pi)$ shape \cite{Li_2015, Li_2016, Li_2017, Li_2018, Park_2019}; the NL method instead uses only the sign-change locus within each redshift bin and is largely insensitive to the overall clustering amplitude. We have shown that the NL yields constraints without directly using BAO or CMB likelihoods, complementary to those from the small-scale shape of $\xi$.

Turning the NL into a robust probe will require characterizing the $\phi$--$c$ likelihood surface with large simulation suites, better-motivated bias priors, a quantified systematics budget, and a consistent framework for combining the NL with BAO and the full-shape measurements (see also Appendix F); none of these appears to present a fundamental obstacle. Whether full-shape analyses on quasi-linear scales already capture the NL information is an open question: the full two-dimensional $\xi$ must formally contain it, but small-scale analyses require numerous nuisance parameters that can dilute cosmological information and introduce prior-volume effects \cite{chudaykin2025priorsscalecutseftbased, holm2023bayesianfrequentistinvestigationprior, Simon_2023}, so isolating the NL as a separate, potentially more robust observable may still be valuable, much as the BAO peak is analyzed separately from the broadband shape. Even if it does not ultimately improve constraints, the NL would remain valuable as an independent consistency test as cosmological data sets become increasingly precise.

\begin{acknowledgments}

\textit{Acknowledgments}---I am grateful to the following agencies for supporting my research: the DOE Office of Science through the grant DE-SC0011840, NASA through ROSES grant 12-EUCLID12-0004 and grant No. 80NSSC24M0021, “Project Infrastructure for the Roman Galaxy Redshift Survey”.

This research used resources of the National Energy Research Scientific Computing Center (NERSC), a Department of Energy User Facility using NERSC award HEP-ERCAP 36297 (desi-2026)

This project was conceived and executed during my summer of 2026 travel. I would like to thank Ofer Lahav (UCL), Alan Heavens (Imperial College), and Florian Beutler (University of Edinburgh) for hosting me at their institutions during this time. I would similarly like to thank the organizers of the ``GGI: Exploring New Frontiers in Cosmology'' conference for the invitation and providing a distraction-free working environment in beautiful Tuscany during the final week of working on this project. 

I thoroughly enjoyed discussing this project with Claude and ChatGPT. I used these systems for writing plotting code and restructuring the code written by me to be more easily readable, that would have otherwise taken much longer. I also acknowledge their help with catching typos, grammar mistakes, and helping improve the flow of the prose after I shared my first draft with them. All scientific ideas and insights, strategic decisions about the direction of project, and the majority of text in the manuscript are otherwise mine, and I take full responsibility for any errors or inaccuracies that may transpire. 

This research used data obtained with the Dark Energy Spectroscopic Instrument (DESI). DESI construction and operations is managed by the Lawrence Berkeley National Laboratory. This material is based upon work supported by the U.S. Department of Energy, Office of Science, Office of High-Energy Physics, under Contract No.~DE--AC02--05CH11231, and by the National Energy Research Scientific Computing Center, a DOE Office of Science User Facility under the same contract. Additional support for DESI was provided by the U.S. National Science Foundation (NSF), Division of Astronomical Sciences under Contract No.~AST-0950945 to the NSF's National Optical-Infrared Astronomy Research Laboratory; the Science and Technology Facilities Council of the United Kingdom; the Gordon and Betty Moore Foundation; the Heising-Simons Foundation; the French Alternative Energies and Atomic Energy Commission (CEA); the National Council of Humanities, Science and Technology of Mexico (CONAHCYT); the Ministry of Science and Innovation of Spain (MICINN); and by the DESI Member Institutions. The DESI Collaboration is honored to be permitted to conduct scientific research on I'oligam Du'ag (Kitt Peak), a mountain with particular significance to the Tohono O'odham Nation. Any opinions, findings, and conclusions or recommendations expressed in this material are those of the authors and do not necessarily reflect the views of the U.S. National Science Foundation, the U.S. Department of Energy, or any of the listed funding agencies.

\end{acknowledgments}

\bibliography{nodalline}

\appendix

\setcounter{equation}{0}
\renewcommand{\theequation}{A\arabic{equation}}

\textit{Appendix A: Amplitude invariance and residual bias dependence}---%
In linear theory, the redshift-space galaxy power spectrum can be written as
\begin{equation}
P_s(k,\mu)
=
b^2\sigma_8^2
\left(1+\beta\mu^2\right)^2
\widehat{P}(k),
\qquad
\beta \equiv \frac{f}{b},
\label{eq:kaiser-amplitude}
\end{equation}
where $\widehat{P}(k)$ denotes the unit-normalized linear power-spectrum shape and $b\sigma_8$ sets the overall clustering amplitude. Because the NL estimator depends on $\xi$ only through $\operatorname{sgn}(\xi)$, it is exactly invariant under any positive rescaling of the overall normalization $b^2\sigma_8^2$ at fixed $\beta$: multiplying $\xi$ by a positive constant leaves the location of its zero contour unchanged. The NL therefore contains no information about $b\sigma_8$ as an overall-amplitude parameter and, unlike a full-shape analysis, requires no modeling of the clustering amplitude or external calibration of $\sigma_8$ at the level of the mean NL prediction.

The residual sensitivity to galaxy bias enters through $\beta=f/b$, which controls the strength of the redshift-space anisotropy and therefore changes the locus at which $\xi$ crosses zero. At fixed cosmology, decreasing the bias increases $\beta$, strengthens the anisotropic distortion, and displaces the NL. This is the dependence explored using the fiducial biases of Appendix D and the independent $10\%$ Gaussian priors adopted in the main analysis.

There is therefore no contradiction between amplitude invariance and the need to marginalize over galaxy bias: the NL is insensitive to $b\sigma_8$ as an overall normalization, but retains a weaker geometric dependence on $\beta$. Since the growth rate is determined by the cosmological model and is approximately
\begin{equation}
f(z) \simeq \Omega_{\rm m}(z)^{0.55}
\end{equation}
as shown in \cite{2005PhRvD..72d3529L}. In flat $\Lambda$CDM, a prior on $b$ at fixed cosmology is equivalent to a prior on $\beta$.

\setcounter{equation}{0}
\renewcommand{\theequation}{B\arabic{equation}}

\textit{Appendix B: Nodal-line estimator}---We parameterize the NL as a straight line in Hesse normal form. The signed distance of bin $i$ from a trial line $(\phi,c)$ is \begin{equation} d_i(\phi,c)=\pi_i\cos\phi-\sigma_i\sin\phi-c , \end{equation} with $\phi$ the angle of the line to the $\sigma$ axis and $c$ its signed perpendicular offset from the origin. We convert this to a softened model sign,
\begin{equation}
m_i(\phi,c)=\tanh\!\left[\frac{d_i(\phi,c)}{\Delta}\right],
\end{equation}
which transitions smoothly between the two sides of the line over the softening scale $\Delta=2\,h^{-1}\,\mathrm{Mpc}$, and fit $(\phi,c)$ by minimizing the summed squared mismatch between the softened model and the measured sign field over the mask $\mathcal{M}$,
\begin{equation}
\mathcal{L}(\phi,c)=\sum_{i\in\mathcal{M}}
\left[q_i-m_i(\phi,c)\right]^2 .
\end{equation}
The minimization is performed first on a grid in $(\phi,c)$ and then refined with a local paraboloid fit about the grid minimum, yielding the best-fit values $(\phi_\star,c_\star)$.

The softening scale $\Delta$ makes $\mathcal{L}$ smooth and suppresses its sensitivity to isolated sign flips near the trial line. As $\Delta\to0$, $m_i\to\operatorname{sgn}(d_i)$ and the loss becomes proportional to the number of masked bins whose signs disagree with the trial line; for an exactly linear NL in the noise-free limit the estimator then recovers the input line to within the grid and local-fit accuracy. 

\textit{Appendix C: Covariance estimation}---We use the 25 mocks per tracer to estimate the measurement covariance of $\phi_\star$ and $c_\star$ under a two-parameter Gaussian approximation. The resulting matrix is block diagonal, with one $2\times2$ block per tracer. Because each block is inverted from an estimate based on $N=25$ mocks, we apply the Hartlap correction $\alpha=(N-p-2)/(N-1)=0.875$ with $p=2$.

\textit{Appendix D: Fiducial tracer biases}---The fiducial tracer biases used in the fixed-bias analysis, and as the centers of the 10\% Gaussian priors in the marginalized analysis, are
$b_\mathrm{LRG1}=1.930$,
$b_\mathrm{LRG2}=2.240$,
$b_\mathrm{ELG1}=1.415$,
$b_\mathrm{ELG2}=1.750$, and
$b_\mathrm{QSO}=1.875$.
These are obtained by profiling the NL likelihood over $\Omega_\mathrm{m}$ and $h$ at the Planck 2018 TT,TE,EE$+$lowE$+$lensing baseline flat-$\Lambda$CDM cosmology \cite{Planck_2020}. These values should be regarded as fiducial calibration values rather than independent external measurements of tracer bias. Because their determination is conditioned on the Planck baseline cosmology, the numerical proof-of-concept constraints are not strictly independent of CMB-conditioned assumptions, although the NL method itself requires no CMB calibration of the sound horizon. A definitive cosmological analysis should instead determine the bias parameters jointly using a validated model or adopt genuinely external bias information. The present results are intended to demonstrate the information content and potential complementarity of the NL rather than to provide final cosmological constraints.

\textit{Appendix E: Direction of the bias degeneracy}---Marginalizing over the tracer biases broadens the cosmological posterior asymmetrically, primarily toward the upper-right region of the $(h,\Omega_\mathrm{m})$ plane, which corresponds to lower bias values. The direction of this extension should not be confused with the motion of the NL in either the correlation-function plane or the fitted-parameter plane. Increasing the bias shifts the NL upward in the $(\sigma,\pi)$ plane and toward the upper right in the fitted $(\phi,c)$ plane. However, because the mapping from $(\Omega_\mathrm{m},h)$ to $(\phi,c)$ is oppositely directed along the relevant degeneracy, this higher-bias displacement corresponds to the lower-left side of the $(h,\Omega_\mathrm{m})$ posterior. The higher-bias direction is more strongly constrained because it shifts the NL toward the positive-$\xi$ region, where $|\xi|$ is larger and its sign is measured more robustly than in the negative-$\xi$ region. Consequently, the posterior develops a longer tail in the cosmological direction corresponding to lower bias.

\textit{Appendix F: Observational systematics}---Because the NL is defined from a restricted feature over an intermediate range of scales, its sensitivity to observational systematics may be no greater than that of established BAO and full-shape measurements, although this must be demonstrated quantitatively. The two dominant effects act at opposite ends of the fitted range: imaging systematics predominantly affect large scales, whereas fiber-assignment incompleteness is most severe at small angular and transverse separations. The NL fit avoids the former through its restricted upper scale and reduces sensitivity to the latter by excluding the smallest total and transverse separations. A full validation against simulations spanning realistic variations in halo occupation, velocity bias, and tracer properties remains necessary.

\end{document}